\documentclass[preprint, showpacs,preprintnumbers,amsmath,amssymb,nofootinbib]{revtex4}
\usepackage{mathrsfs}
\usepackage{amsmath}
\usepackage{amsfonts}
\usepackage{latexsym}
\usepackage{amsfonts}
\usepackage{graphicx}
\usepackage{epsf}
\usepackage{dcolumn}
\usepackage{bm}
\newcommand{\bea}[1]{\begin{eqnarray}\label{#1}}
 \newcommand{\eea}{\end{eqnarray}}

 \def\gsim{ \lower .75ex \hbox{$\sim$} \llap{\raise .27ex \hbox{$>$}} }
 \def\lsim{ \lower .75ex \hbox{$\sim$} \llap{\raise .27ex \hbox{$<$}} }
\def\/{\over}

\begin{document}

\title{\bf Oscillating universe in the DGP braneworld}

\author{ Kaituo Zhang$^1$, Puxun Wu$^2$, Hongwei Yu$^{1,2}$ \footnote{Corresponding
author:hwyu@hunnu.edu.cn}}
\address{ $^1$Department of Physics and Key Laboratory of Low Dimensional
Quantum Structures and Quantum Control of Ministry of Education,
Hunan Normal University, Changsha, Hunan 410081, China\\
$^2$ Center for Nonlinear Science and Department of Physics, Ningbo
University,  Ningbo, Zhejiang 315211, China}

\begin{abstract}
With a method in which the Friedmann equation is written in a form
such that evolution of  the scale factor can be treated  as that of
a particle in a ``potential", we classify all possible cosmic
evolutions in the DGP braneworld scenario with the dark radiation
term retained. By assuming that the energy component is pressureless
matter, radiation or vacuum energy, respectively, we find that in
the matter or vacuum energy dominated case, the scale factor has a
minimum value $a_0$. In the matter dominated case,  the big bang
singularity can be avoided in some special circumstances, and there
may exist an oscillating universe or a bouncing one. If the cosmic
scale factor is in the oscillating region initially, the universe
may undergo an oscillation. After a number of oscillations, it may
evolve to the bounce point through quantum tunneling and then
expand.  However, if the universe contracts initially from an
infinite scale, it can turn around and then expand forever. In the
vacuum energy dominated case, there exists a stable Einstein static
state to avoid the big bang singularity.   However, in certain
circumstances in the matter or vacuum energy dominated case,  a new
kind of singularity may occur at $a_0$ as a result of the
discontinuity of the scale factor. In the radiation dominated case,
the universe may originate from the big bang singularity, but a
bouncing universe which avoids this singularity is also possible.
\end{abstract}


\maketitle
\section{Introduction}
The modified gravity has spurred  an increasing deal of interest
recently, because it can explain, without the introduction of an
exotic dark energy, the present  accelerating cosmic expansion
discovered firstly from the Type Ia Supernovae (Sne
Ia)~\cite{Perlmutter1999, Riess1998, Riess2004, Riess2006,
Astier2006, Wood2007}. Among the modified gravity theories, the
Dvali-Gabadadze-Porrati (DGP) braneworld scenario~\cite{Dvali2000},
generalized firstly to cosmology by Deffayet~\cite{Deffayet2001}, is
a very simple and popular one. The DGP theory starts with the idea
that our observed four-dimensional Universe resides in a
five-dimensional, infinite-volume Minkowski bulk and the whole
energy-momentum is confined on a three dimensional spacial brane. In
contrast to the Randall-Sundrum~\cite{Randall1999} and
Shtanov-Sahni~\cite{Shtanov2003} braneworld scenarios with high
energy modifications to general relativity, the DGP brane produces a
low energy modification (for a review of the phenomenology of the
DGP model, see Ref.~\cite{Lue2006}).

Since there are two different ways to embed the 4-dimensional brane
universe into the 5-dimensional spacetime, the DGP model has two
separate branches denoted by $\epsilon=\pm1$.  The $\epsilon=+1$
branch is  self-accelerating in the sense that  the universe is
rendered to accelerate at late times due to the  lowly leaking of
the gravity off our four-dimensional world into an extra dimension
and to evolve eventually into a de Sitter phase~\cite{Deffayet2001}.
However,  the $\epsilon=-1$ branch is very different since it does
not self-accelerate. Thus, in order to explain the present cosmic
acceleration in this branch, dark energy is  required on the brane,
like in the  LDGP model~\cite{Lue2004} and QDGP
model~\cite{Chimento2006}.

The inflation and preheating on the DGP brane have been discussed in
Refs.~\cite{Bouhmadi-Lopez2004, Cai2004, Papantonopoulos2004,
Zhang2004, Zhang2006, Campo2007} and  some new characteristics have
been found. For example, the  DGP inflation driven by a single
scalar field with an exponential potential yields much better
consistency with the current observation data~\cite{Cai2004}.
Recently, we discussed the stability of the Einstein static universe
in the DGP scenario~\cite{Zhang} and obtained that the universe can
stay at this stable Einstein static state past-eternally, undergo a
series of infinite, non-singular oscillations, and then evolve to
inflation. Therefore, the big bang singularity  can be avoided.
Moreover, the cosmic background evolutions in the DGP model have
been studied in ~\cite{Shtanov2003, Sahni2003, Sahni2005}. With a
large value of the dark radiation term, it was found that the
spatially flat DGP braneworld gives the same dynamical possibilities
of the cosmic evolution as a closed FRW universe and these
possibilities include  the oscillating, the bouncing, the Einstein
static universes  and the so-called loitering universe (see Fig.(4)
of \cite{Sahni2005}). In the present paper, we plan to classify all
possible cosmic evolutions in the DGP braneworld with a method in
which the Friedmann equation is written in a form such that
evolution of  the scale factor can be treated as that of  a particle
in a ``potential". The effect of the dark radiation are also
considered in contrast to Ref.~\cite{Zhang}.  Different from Ref.
\cite{Sahni2005}, we keep, in our discussion,  the spatial curvature
term and do not impose  the condition of a large value of the dark
radiation. Let us note that this method has been used to classify
the cosmic evolution in the Horava-Lifshitz
gravity~\cite{Maeda2010}.

\section{The Friedmann equation in the DGP braneworld}

We consider a homogeneous and isotropic universe
described by the Friedmann-Robertson-Walker (FRW) metric
\begin{eqnarray}
ds^2=-dt^2+a^2(t)\bigg(\frac{dr^2}{1-kr^2}+r^2d^2\Omega\bigg)\;,
\end{eqnarray}
where $a$ is the cosmic scale factor and $k$ is the constant
curvature of the three-space of the FRW metric. In the DGP brane
scenario, the Friedmann equation can be written as~\cite{Maeda2003}
\begin{eqnarray}
H^2+\frac{k}{a^2}=\frac{1}{3\mu^2}[\rho+\rho_0(1+\epsilon
\mathcal{A}(\rho,a))]\;,
\end{eqnarray}
where $H$ is the Hubble parameter, $\rho$ the total energy density
and  $\mu$ a  parameter denoting the strength of the induced gravity
on the brane. $\mathcal{A}$ is given by
\begin{eqnarray}
\mathcal{A}=\bigg[\mathcal{A}_0^2+\frac{2\eta}{\rho_0}\bigg(\rho-\mu^2\frac{\mathcal
{E}_0}{a^4}\bigg)\bigg]^{1/2}\;,
\end{eqnarray}
where \begin{eqnarray}\mathcal{A}_0=\sqrt{1-2\eta
\frac{\mu^2\Lambda}{\rho_0}},\quad \eta=\frac{6m_5^6}{\rho_0\mu^2}
\;\;\; (0<\eta\leq 1), \quad
\rho_0=m_\lambda^4+6\frac{m_5^6}{\mu^2}\;,\end{eqnarray} with
$\Lambda$ defined as
\begin{eqnarray}\Lambda=\frac{1}{2}(^{(5)}\Lambda+\frac{1}{6}\kappa_5^4\lambda^2)\;.
\end{eqnarray}
Here $\kappa_5$ is the 5-dimensional Newton constant,
$^{(5)}\Lambda$  the 5-dimensional cosmological constant in the
bulk, $\lambda$  the brane tension, and $\mathcal {E}_0$   an
integration constant related to the Weyl radiation (dark radiation)
which is assumed to be positive in this paper. For simplicity, we
restrict ourselves to the Randall- Sundrum critical case, i.e.
$\Lambda=0$, then Eq.(2) simplifies to
\begin{eqnarray}
H^2+\frac{k}{a^2}=\frac{1}{3\mu^2}\bigg(\rho+\rho_0+\epsilon\rho_0\sqrt{1+\frac{2\eta}{\rho_0}\bigg(\rho-\mu^2\frac{\mathcal
{E}_0}{a^4}\bigg)}\;\bigg).
\end{eqnarray}
In the very early era of the universe the total energy
density should be very high.  Thus, we will, in the following, only
consider the ultra high energy limit, $\rho\gg\rho_0$. %$\gg$$m_\lambda^4$.
In addition, we let $\eta=1$. As a result, the Friedmann equation reduces to
\begin{eqnarray}\label{FEq}
H^2+\frac{k}{a^2}=\frac{1}{3\mu^2}\bigg(\rho+\epsilon\sqrt{2\rho_0\bigg(\rho-\mu^2\frac{\mathcal
{E}_0}{a^4}\bigg)}\;\bigg).
\end{eqnarray}
 The dark radiation term is retained here in contrast to Ref~~\cite{Zhang} where it is
 neglected. When $a$ is small this
term is very important. It is easy to see that the above equation
describes a 4-dimensional gravity with minor corrections, which
implies that $\mu$ must have an energy scale as the Planck one in
the DGP model.

For the cosmic energy, we assume that it has a constant equation of state
$\omega$ and thus its density  can be expressed as
\begin{eqnarray}
\rho=\frac{g}{a^{3(1+\omega)}},
\end{eqnarray}
where $g$ is a constant. In the following, we
take $\omega=-1, 1/3$ or $0$, which corresponds to  the vacuum energy, radiation, or pressureless matter dominated universe,  respectively.
 Thus the Friedmann equation becomes
\begin{eqnarray}
H^2+\frac{k}{a^2}=\frac{1}{3\mu^2}\bigg(\frac{g}{a^{3(1+\omega)}}+\epsilon\sqrt{2\rho_0\bigg(\frac{g}{a^{3(1+\omega)}}-\mu^2\frac{\mathcal{E}_0}{a^4}\bigg)}\ \bigg).
\end{eqnarray}
Clearly, $a\geq a_0=(\frac{\mu^2\mathcal{E}_0}{g})^{1/(1-3\omega)}$
is required when $\omega\neq 1/3$, which means that  the universe
begins to evolve at $a\geq a_0$ rather than $a=0$. So, the classical
big bang singularity can be avoided. Let us note that this finite
size initial universe can be created from "nothing" through quantum
tunneling~\cite{Hartle:1983ai,Vilenkin:1984wp}.  For $\omega=1/3$,
$g\geq \mu^2\mathcal{E}_0$ is needed and in this case the universe
can originate from $a=0$.

Now we rewrite the Friedmann equation in the following form
\begin{eqnarray}
\dot{a}^2+V(a)=0,
\end{eqnarray}
where
\begin{eqnarray}\label{va}
V(a)=k-\frac{1}{3\mu^2}\frac{g}{a^{3\omega+1}}-\frac{\epsilon}{3\mu^2}\sqrt{2\rho_0\bigg(\frac{g}{a^{3\omega-1}}-\mu^2\mathcal{E}_0\bigg)}\;.
\end{eqnarray}
Thus $V$ can be regarded as a ``potential" and the scale factor $a$
changes as a particle moving in it.  This ``potential" must satisfy
the condition  $V(a)\leq0$. This gives the possible range of $a$
when the universe evolves. Therefore, we can classify the types of
the universe by the signs of $k$ and $\epsilon$, and by the values
of other parameters.

All cosmic evolution types  in the DGP braneworld are:

(1) [Bounce]: If $V(a)\leq0$ for $a\in[a_T,\infty)$ and the equality
holds at $a=a_T$, a spacetime initially contracts from an infinite
scale, and it eventually turns around at the finite scale $a_T$, and
then expands forever;

(2) [Oscillation]:  $V(a)\leq0$ for $a\in[a_{min},a_{max}]$ and the
equality occurs at $a=a_{min}$ and $a=a_{max}$, a spacetime
oscillates between two finite scale factors;

(3) [$FS\Rightarrow \infty$]: $V(a)<0$ for $a\in[a_0,\infty)$. The
universe starts at finite size(FS) $a_0$ and expands forever.

(4) [$BB\Rightarrow BC$]: $V(a)\leq0$ for $a\in(0,a_T]$ and the equality holds at $a=a_T$. A spacetime starts from a big bang (BB) and expands. It turns around at $a=a_T$ and then contracts. Eventually, the universe contracts to a big crunch (BC). $a_T$ is the scale factor where the universe turns around from expansion to contraction.

(5) [$BB\Rightarrow \infty$ or $\infty\Rightarrow BC$]: $V(a)<0$ for any positive values of $a$, a spacetime starts from a big bang and expands forever, or the spacetime always contracts to a big crunch.

(6) [$FS\Rightarrow FS$]: $V(a)\leq0$ for $a\in[a_0,a_m]$ and the
equality holds only at $a=a_m$. A spacetime starts from a finite
scale $a_0$ and expands. It turns around at $a=a_m$ and begins to
contract. When the universe contracts to the minimum scale $a=a_0$,
it should expand again. However,  its evolution will be
discontinuous at $a=a_0$ since the potential $V(a_0)\neq0$. Thus,
there exists a new singularity in this type.

\section{The evolution of a matter-dominated universe in the DGP braneworld}

If the universe is dominated by pressureless matter $(\omega=0)$, the cosmic energy density can be expressed as $\rho=\frac{g_m}{a^3}$. Thus, the potential becomes
\begin{eqnarray}\label{vaa}
V(a)=k-\frac{1}{3\mu^2}\frac{g_m}{a}-\frac{\epsilon}{3\mu^2}\sqrt{2\rho_0(g_m
a-\mu^2\mathcal{E}_0)}\;.
\end{eqnarray}
Clearly, $ a_0=\frac{\mu^2\mathcal{E}_0}{g_m}$ and
$H(a_0)\neq 0$ in general except for the case
\begin{eqnarray}\label{b1} g_m^2=3k\mu^4\mathcal{E}_0\;.\end{eqnarray}
 This condition gives a boundary to obtain an oscillating
universe.

A static  universe appears if there is a solution $a=a_S\geq a_0$
which satisfies $V(a_S)=0$ and $V'(a_S)=0$. At $a_S$, both the
cosmic expansion speed and acceleration equal to zero and thus the
universe can stay at this point if it is stable. Differentiating
$V(a)$ with respect to $a$, we have
\begin{eqnarray}
V'(a)=\frac{1}{3\mu^2}\frac{g_m}{a^2}-\frac{\epsilon\sqrt{\rho_0}}{3\sqrt{2}\mu^2}\frac{g_m}{\sqrt{g_m
a-\mu^2\mathcal{E}_0}}\;.
\end{eqnarray}
Combining $V(a)=0$ and $V'(a)=0$, we obtain, to get a static
universe,  a relation between $g_m$ and other parameters:
\begin{eqnarray}\label{gm}
g_m=g_m^{\pm}=\sqrt{\frac{\mu^3}{9\rho_0}\bigg(9k\mu(\mu^2+2\mathcal
{E}_0\rho_0)\pm\sqrt{3}(3\mu^2-2\mathcal{E}_0\rho_0)^\frac{3}{2}\bigg)}\;,
\end{eqnarray}
which gives another two boundaries  for obtaining the oscillating
universe. Now, we have three boundary conditions (Eq.~(\ref{b1},
\ref{gm})) for an oscillation.  Using Eq.~(\ref{gm}) and
Eq.~(\ref{vaa}), one can find the static state solution
\begin{eqnarray}\label{as}
a_S=a_S^{\pm}=\frac{\sqrt{\mu}(3\mu^2+2\mathcal{E}_0\rho_0\pm k\sqrt{9\mu^4-6\mathcal{E}_0\rho_0\mu^2})}{\sqrt{\rho_0}\sqrt{9k\mu(\mu^2+2\mathcal{E}_0\rho_0)\pm\sqrt{3}(3\mu^2-2\mathcal{E}_0\rho_0)^\frac{3}{2}}}\;,
\end{eqnarray}
which is a double root of the equation $V=0$ under the condition
$V'=0$. Then the third root is easy to find
\begin{eqnarray}\label{at}
a_T=a_T^{\pm}=\frac{\sqrt{\mu}(15\mu^2-2\mathcal{E}_0\rho_0\mp 4k\sqrt{9\mu^4-6\mathcal{E}_0\rho_0\mu^2})}{2\sqrt{\rho_0}
\sqrt{9k\mu(\mu^2+2\mathcal{E}_0\rho_0)\pm\sqrt{3}(3\mu^2-2\mathcal{E}_0\rho_0)^\frac{3}{2}}}\;.
\end{eqnarray}
It corresponds to the radius where the universe turns around or
bounces.

Now we divide our discussion into two cases: $\epsilon=+1$, and $\epsilon=-1$.

\subsection{$\epsilon=+1$}

Since the oscillating universe exists only in the case of
$\epsilon=+1$ and $k=1$, we first focus on this case.

\subsubsection{$k=1$}

\begin{figure}[htbp]
\includegraphics[width=7cm]{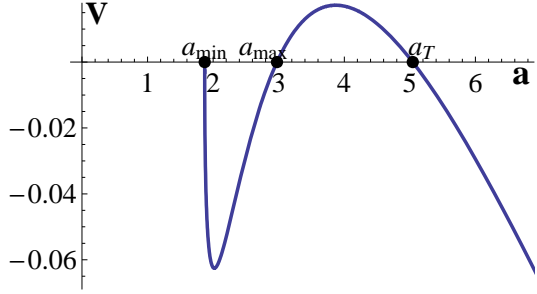}
\caption{\label{F1} The potential ${V}(a)$ for a matter dominated
universe under conditions $g_m^2=3\mu^4\mathcal{E}_0$ and
$\frac{\mathcal{E}_0\rho_0}{\mu^2}<\frac{9}{8}$. The constants are
set as $\mu=1$, $\rho_0=0.1$, $g_m=5.61249$, and
$\mathcal{E}_0=10.5$. The radii of the oscillation are
$a_{min}=1.87083$ and $a_{max}=2.97382$. The period of this
oscillation is $T=18.3324$. The largest root $a_T=5.04402$
corresponds to a turning radius of a bouncing universe. }
\end{figure}

\begin{figure}[htbp]
\includegraphics[width=7cm]{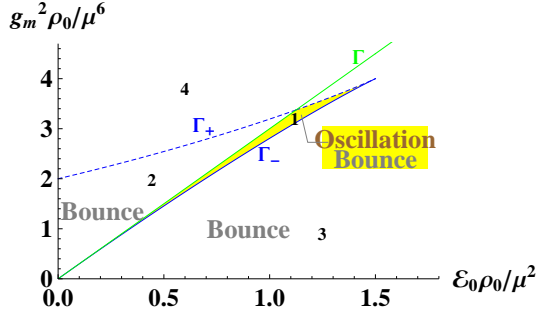}
 \caption{\label{F2} Phase diagram of spacetimes in  $(\frac{g_m^2\rho_0}{\mu^6},
\frac{\mathcal{E}_0\rho_0}{\mu^2})$ plane for a matter dominated universe. An oscillating universe is
found in Region 1. A bounce one is found in Regions 1, 2, and
3. The unstable and stable static universes exist on the boundaries
$\Gamma_+$ and $\Gamma_-$, respectively. }
\end{figure}

By introducing $\tilde{a}=\sqrt{g_m a-\mu^2\mathcal {E}_0}$,
Eq.~(\ref{va}) becomes
\begin{eqnarray}\label{Vt}
V(\tilde{a})=-\frac{\sqrt{2\rho_0}}{3\mu^2}\frac{g_m}{(\tilde{a}^2+\mu^2\mathcal{E}_0)}
\bigg(\tilde{a}^3-\frac{3\mu^2}{\epsilon\sqrt{2\rho_0}}\tilde{a}^2+\mathcal{E}_0\mu^2\tilde{a}+\frac{g_m^2
-3\mu^4\mathcal{E}_0}{\epsilon\sqrt{2\rho_0}}\bigg)\;.
\end{eqnarray}
Apparently, when  $V=0$, we get  a cubic equation of $\tilde{a}$,
which can be expressed as
\begin{eqnarray}
-\frac{\sqrt{2\rho_0}}{3\mu^2}\frac{g_m}{(\tilde{a}^2+\mu^2\mathcal{E}_0)}
(\tilde{a}-\tilde{a}_{min})(\tilde{a}-\tilde{a}_{max})(\tilde{a}-\tilde{a}_T)=0\;,
\end{eqnarray}
with $\tilde{a}_{min}$, $\tilde{a}_{max}$ and $\tilde{a}_T$ being
three solutions. Assuming
$0\leq\tilde{a}_{min}\leq\tilde{a}_{max}\leq\tilde{a}_T$, if
$V(\tilde{a})\leq0$ in $\tilde{a}\in[\tilde{a}_{min},
\tilde{a}_{max}]$ and the equality holds when
$\tilde{a}=\tilde{a}_{min}$ and $\tilde{a}=\tilde{a}_{max}$, the
universe oscillates between two finite scales; if $V(\tilde{a})\leq
0$ in $\tilde{a}\in[\tilde{a}_T, \infty)$ and the equality holds
when $\tilde{a}=\tilde{a}_T$, it corresponds to a bounce scenario
and the universe bounces at $\tilde{a}_T$. For a simple example, let
$g_m^2=3\mu^4\mathcal{E}_0$ (the boundary $\Gamma$ (Eq.~(\ref{b1}))
in Fig.~(\ref{F2}) for obtaining the oscillating universe)  in the
potential (Eq.~(\ref{Vt})) , we have
\begin{eqnarray}
a_{min}=a_0\;,
\end{eqnarray}
\begin{eqnarray}
a_{max}=\frac{\sqrt{3}({3\mu^2-\sqrt{9\mu^4-8\mathcal{E}_0\rho_0\mu^2}}}{4\rho_0\sqrt{\mathcal{E}_0}})\;,
\end{eqnarray}
\begin{eqnarray}
a_T=\frac{\sqrt{3}({3\mu^2+\sqrt{9\mu^4-8\mathcal{E}_0\rho_0\mu^2}}}{4\rho_0\sqrt{\mathcal{E}_0}})\;.
\end{eqnarray}
For the case $\frac{\mathcal{E}_0\rho_0}{\mu^2}<\frac{9}{8}$, we
have $0<a_{min}<a_{max}<a_T$. In Fig.~(\ref{F1}), we plot the
evolutionary curve of $V(a)$. It is easy to see that $V(a)\leq 0$ in
$a\in[a_{min}, a_{max}]$ and $a\in[a_T, \infty)$, which means there
is an oscillating universe ($a\in[a_{min}, a_{max}]$) or a bouncing
one ($a\in[a_T,\infty)$). Thus, if the universe is in  the region
$[a_{min}, a_{max}]$ initially, it  may undergo an oscillation.
After a number of oscillations, it may evolve to the bounce point
$a_T$ through quantum tunneling.  If the universe contracts
initially from an infinite scale, it can turn around at $a_T$ and
then expand forever.

For a general case, we find that there is an oscillating universe if
the following conditions are satisfied
\begin{eqnarray}\label{ec}
 \mathcal{E}_0\rho_0<\frac{9}{8}\mu^2 \;, \quad\quad&g_m^-<g_m\leq\sqrt{3\mathcal{E}_0}\mu^2
  \nonumber \\
 \frac{9}{8}\mu^2\leq\mathcal{E}_0\rho_0<\frac{3}{2}\mu^2 \;\;,\;&g_m^-<g_m<g_m^+,
\end{eqnarray}
where $g_m^\pm$ is defined in Eq.~(\ref{gm}). Using above equations,
we obtain  the allowed region  in $(\frac{g_m^2\rho_0}{\mu^6},
\frac{\mathcal{E}_0\rho_0}{\mu^2})$ plane (Fig.~(\ref{F2})) for an
oscillating universe or a bouncing one. The boundaries curves
$\Gamma_\pm$ are defined as $g_m=g_m^\pm$  (Eq.~(\ref{gm})) and
curve $\Gamma$ as $g_m=\sqrt{3\mathcal{E}_0}\mu^2$ (Eq.~(\ref{b1})).

\begin{figure}[htbp]
\includegraphics[width=7cm]{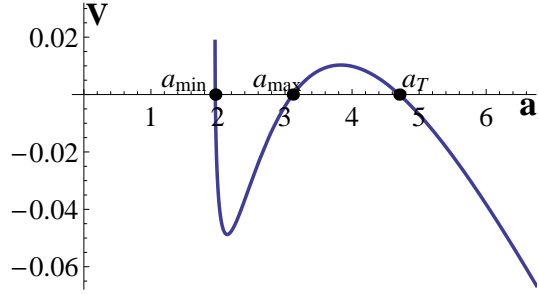}
\caption{\label{F3} The potential ${V}(a)$ for an oscillating
universe or a bouncing one in Region 1 of Fig.~(\ref{F2}).  The
constants are chosen as $\mu=1$, $\rho_0=0.1$, $g_m=5.75$, and
$\mathcal{E}_0=11.25$. The radii  are $a_{min}=1.96042$,
$a_{max}=3.11647$ and $a_T=4.70572$. The period of an oscillation is
$T=21.9044$.}
\end{figure}

The period of an oscillation  can be calculated through
\begin{eqnarray}
T:=2\int_{a_{min}}^{a_{max}} \frac{d a}{\sqrt{-V(a)}}
\end{eqnarray}
where $a_{max}$ and $a_{min}$ are the maximum and
minimum radius of the oscillating universe.

In Fig.~(\ref{F3}), we give the evolutionary curve of the potential
with the model parameters satisfying Eq.~(\ref{ec}). From this
figure, we find that there is an oscillating universe between
$a_{min}$ and $a_{max}$, or a bouncing one in $[a_T, \infty)$.
Therefore, a similar cosmic evolution as shown in Fig.~(\ref{F1}) is
obtained.

\begin{figure}[htbp]
\includegraphics[width=7cm]{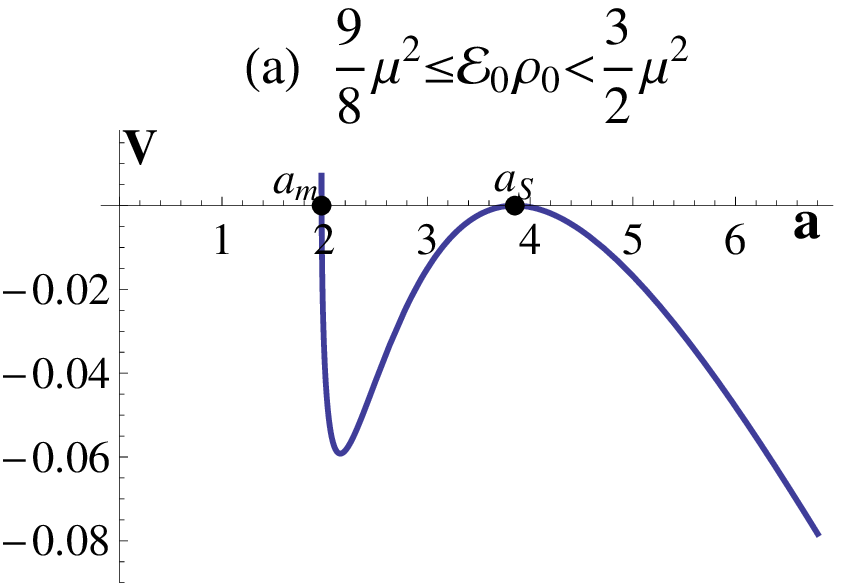}\quad\includegraphics[width=7cm]{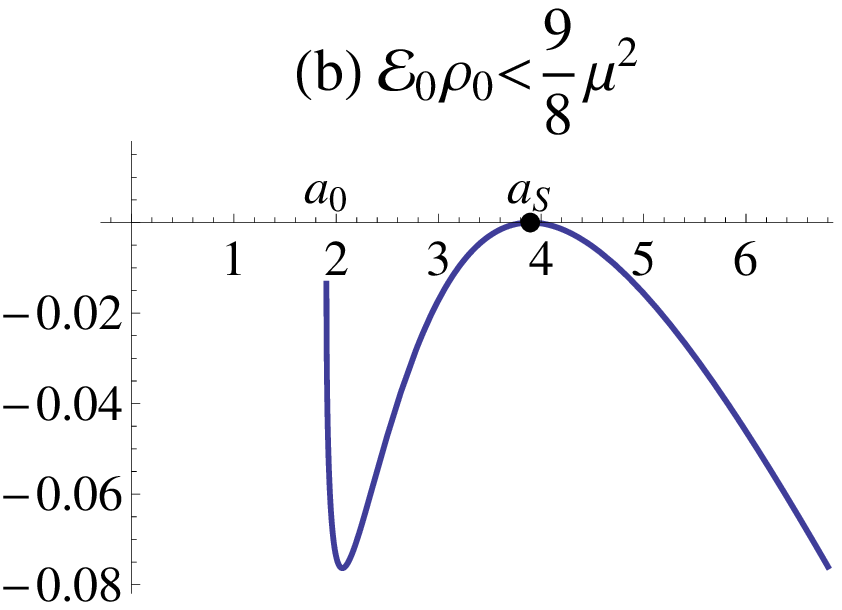}
\caption{\label{F6} The potential ${V}(a)$ in a matter dominated universe for an unstable static
universe (on boundary $\Gamma_+$ of Fig.~(\ref{F2})). The constants are chosen
$\mu=1$, $\rho_0=0.1$, $\mathcal{E}_0=11.5$, $g_m=5.84184$ (left), and $\mu=1$, $\rho_0=0.1$, $\mathcal{E}_0=11$, $g_m=5.77729$ (right). The radii are $a_m=1.96954$, $a_S=3.85103$ (left) and $a_0=1.90400$, $a_S=3.89410$ (right). }
\end{figure}

On the $\Gamma_+$ curve,  the unstable static universe appears.
 The solution
is
\begin{eqnarray}
a_S=a_S^+\;,
\end{eqnarray}
with $a_s^+$ given in Eq.~(\ref{as}), which
is a double solution of $V(a)=0$. If $\frac{9}{8}\mu^2\leq\mathcal{E}_0\rho_0<\frac{3}{2}\mu^2$, the third root $a_m$ is
\begin{eqnarray}
a_m=a_T^-\;,
\end{eqnarray}
with $a_T^-$ given in Eq.~(\ref{at}). In Fig.~(\ref{F6}) we plot the
evolution of the potential. When $a=a_S$, both $V$ and $V'$ vanish,
but this $a_S$ solution is unstable. Therefore, in the
$\frac{9}{8}\mu^2\leq\mathcal{E}_0\rho_0<\frac{3}{2}\mu^2$ case ,
the universe can oscillate between $a_m$ and $a_S$, and it can also
evolve directly from $a_m$ to $\infty$ or evolve to  $\infty$ after
some oscillations with no need of quantum tunneling to make it
happen. If the universe contracts initially from an infinite scale,
it can turn around at $a_S$, or  pass  through  $a_S$ and  bounce at
$a_m$, then oscillate between $a_m$ and $a_S$. For the
$\mathcal{E}_0\rho_0<\frac{9}{8}\mu^2$  case, if the universe
initially evolves from $a_0$, it can expand to $a_{S}$,  and then it
can further expand to infinity or  turn around. Once the universe
bounces at $a_S$ and contracts to $a_0$, there will  appear a new
singularity since at $a_0$ the Hubble parameter $H(a_0)$ is nonzero.
That is, when the universe contracts to $a_{0}$ and then expands, it
has to  evolve discontinuously at $a_0$.

\begin{figure}[htbp]
\includegraphics[width=7cm]{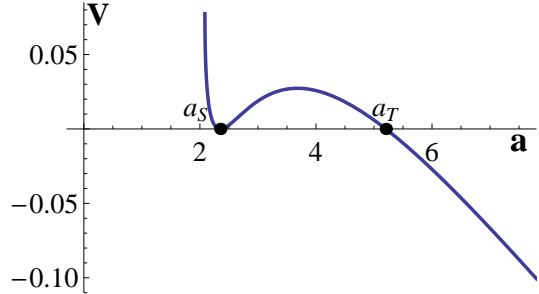}
\caption{\label{F7} The potential $V(a)$  in a matter dominated
universe for a stable static universe and a bouncing one (the
($\Gamma_-$) curve of Fig.~(\ref{F2})). The constants are chosen as
$\mu=1$, $\rho_0=0.1$, $g_m=5.75374$, and $\mathcal{E}_0=12$. The
radius of stable static universe is $a_S=2.35114$ and the bounce
radius  is $a_T=5.20431$.}
\end{figure}

A stable static universe or a bouncing one exists on the $\Gamma_-$
curve. The stable static  solution is
\begin{eqnarray}
a_S=a_S^-
\end{eqnarray}
with $a_s^-$ given in Eq.~(\ref{as}), while the turning radius $a_T$
of a bouncing universe is given by
\begin{eqnarray}
a_T=a_T^+
\end{eqnarray}
with $a_T^+$ given in Eq.~(\ref{at}). Fig.~(\ref{F7}) gives the
potential with model parameter in the $\Gamma_-$ curve. There are
two solutions ($a_S$, $a_T$) for $V=0$.  At $a=a_S$, both $V$ and
$V'$ are equal to zero and apparently $a=a_S$ corresponds to a
stable solution. Thus, the universe can stay at this finite radius
past-eternally. When $a\geq a_T$, $V\leq0$, which corresponds to a
bouncing universe. If the universe stays at $a_s$ initially, after a
long time, it can quantum mechanically tunnel to the bounce point
$a_T$ and then expand. If the universe contracts initially from an
infinite scale, it will turn around at $a_T$.

\begin{figure}[htbp]
\includegraphics[width=7cm]{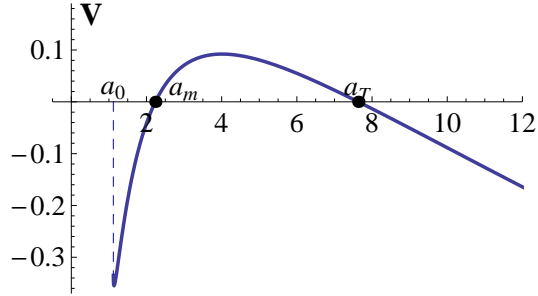}
\caption{\label{F4} The potential ${V}(a)$  in a matter dominated universe with model parameters
in Region 2  of Fig.~(\ref{F2}) . The constants
are chosen as $\mu=1$, $\rho_0=0.1$, $g_m=4.47214$, and $\mathcal{E}_0=5$. The radii are $a_0=1.11803$, $a_m=2.23607$, and $a_T=7.63441$.}
\end{figure}

Fig.~(\ref{F4}) shows the evolution of the potential ${V}(a)$ with
the model parameters in  Region 2 of Fig.~(\ref{F2}). From this
figure, we find that a bouncing universe is obtained since $V\leq0$
in $a\in [a_T, \infty)$. In addition, $V\leq 0$ in $a\in [a_0,
a_{m}]$, but $V=0$ occurs only at $a=a_{m}$. Thus, if the universe
turns around at $a_{m}$ and contracts to $a_0$, as shown in the left
panel of Fig.~(\ref{F6}), there appears a  singularity at $a_0$.  Of
course, if the universe evolves from $a_0$ to $a_{m}$ and then
quantum tunnels to $a_T$ directly, this singularity can be avoided.

\begin{figure}[htbp]
\includegraphics[width=7cm]{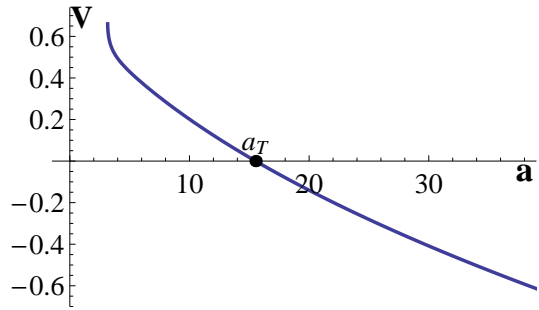}
\caption{\label{F5} The potential ${V}(a)$ in a matter dominated
universe for  a bouncing universe (Region 3 of Fig.~(\ref{F2})). The
constants are $\mu=1$, $\rho_0=0.1$, $g_m=3.16228$, and
$\mathcal{E}_0=10$. The radius is $a_T=15.5259$.}
\end{figure}

Fig.~(\ref{F5}) shows the evolution of the potential ${V}(a)$ with
the model parameters in  Region 3 of Fig.~(\ref{F2}). Apparently,  a
bouncing universe is obtained.

\begin{figure}[htbp]
\includegraphics[width=7cm]{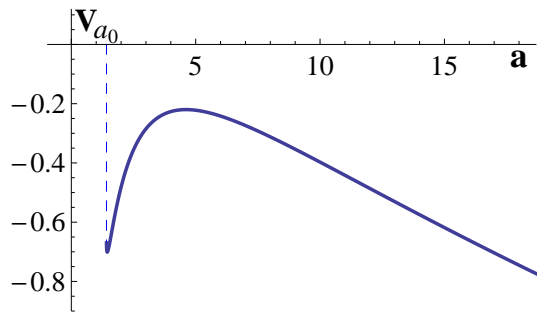}
\caption{\label{F8} The potential ${V}(a)$ in a matter dominated universe for model parameters in
Region 4 of Fig.~(\ref{F2}). The constants are
chosen as $\mu=1$, $\rho_0=0.1$, $g_m=7.07107$, and $\mathcal{E}_0=10$.}
\end{figure}

We plot Fig.~(\ref{F8}) to give the evolution of the potential with
the model parameters in  Region 4 of Fig.~(\ref{F2}). In this case
$V<0$ for $a\geq a_0$. Thus an $FS\Rightarrow \infty$ type is
obtained.

\subsubsection{$k=-1$}

\begin{figure}[htbp]
\includegraphics[width=7cm]{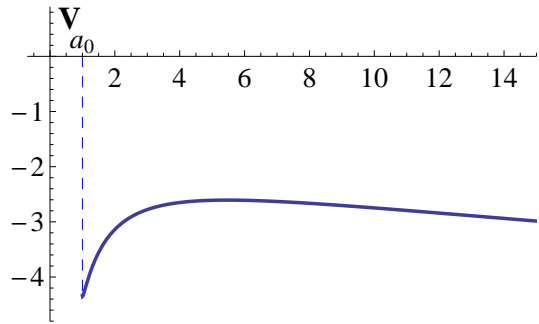}
 \caption{\label{F9} The potential $V(a)$ in a matter dominated universe for the case $\epsilon =+1$, and $k=-1$. The spacetime will always expand. The constants are chosen as $\mu=1$, $\rho_0=0.1$, $g_m=10$ and $\mathcal{E}_0=10$.}
\end{figure}

In this case, the potential is given by
\begin{eqnarray}
V(a)=-1-\frac{1}{3\mu^2}\frac{g_m}{a}-\frac{\sqrt{2\rho_0}}{3\mu^2}\sqrt{g_m a-\mu^2\mathcal{E}_0}\;.
\end{eqnarray}
Fig.~(\ref{F9}) shows the evolution of this potential, from which
one can see that, as Fig.~(\ref{F8}), the potential $V(a)$ is always
negative for  $a\geq a_0$, which means that the cosmic evolution
type is $FS\Rightarrow \infty$.

\subsection{$\epsilon =-1$}

In this case, we get
\begin{eqnarray}
V'(a)=\frac{1}{3\mu^2}\frac{g_m}{a^2}+\frac{\sqrt{\rho_0}}{3\sqrt{2}\mu^2}\frac{g_m}{\sqrt{g_m a-\mu^2\mathcal{E}_0}}.
\end{eqnarray}
Since $V'(a)$ is always positive, the potential is an increasing function.  Therefore, the cosmic evolution will be simpler than the case of $\epsilon=+1$.

\subsubsection{$k=1$}

\begin{figure}[htbp]
\includegraphics[width=7cm]{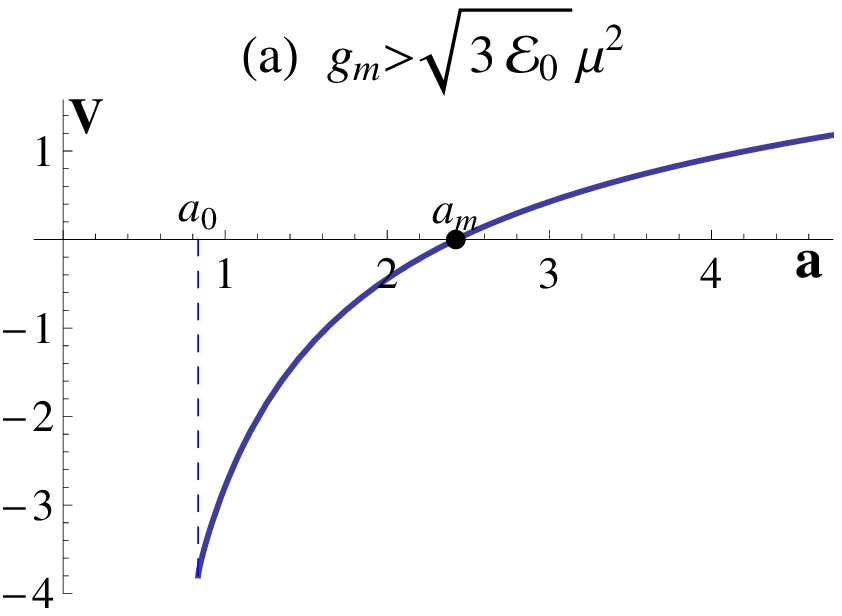}\quad\includegraphics[width=7cm]{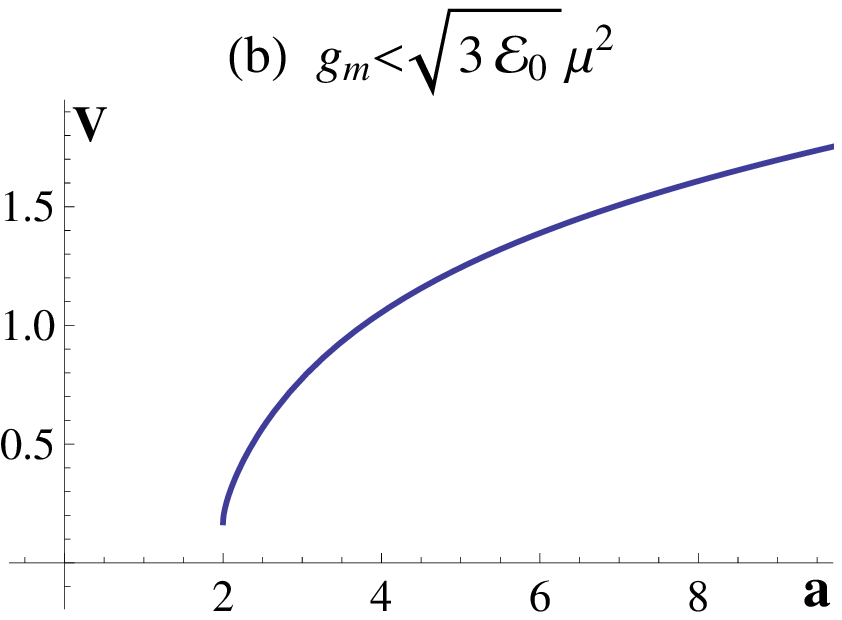}
\caption{\label{F10} The potential ${V}(a)$ in a matter dominated universe  for $\epsilon=-1$ and $k=1$ with
$g_m>\sqrt{3\mathcal{E}_0}\mu^2$ (left) and $g_m<\sqrt{3\mathcal{E}_0}\mu^2$ (right). The constants are chosen as $\mu=1$, $\rho_0=0.1$, $\mathcal{E}_0=10$, $g_m=12$ (left) and $\mu=1$, $\rho_0=0.1$, $\mathcal{E}_0=10$, $g_m=5$ (right). }
\end{figure}

From Fig.~(\ref{F10}) one can see there is no solution for $V\leq0$
if $g_m<\sqrt{3\mathcal{E}_0}\mu^2$. When
$g_m>\sqrt{3\mathcal{E}_0}\mu^2$, if $a_{0}\leq a\leq a_m$,
$V\leq0$, while $V=0$ occurs only at $a_m$. So, the universe can
evolve between $a_0$ and $a_m$, but there is a singularity at $a_0$
since $H(a_0)\neq 0$ at this point, which means that this cosmic
evolution type is $FS\Rightarrow FS$. When
$g_m=\sqrt{3\mathcal{E}_0}\mu^2$, there is only one point $a=a_0$
for $V\leq0$.

\subsubsection{$k=-1$}

\begin{figure}[htbp]
\includegraphics[width=7cm]{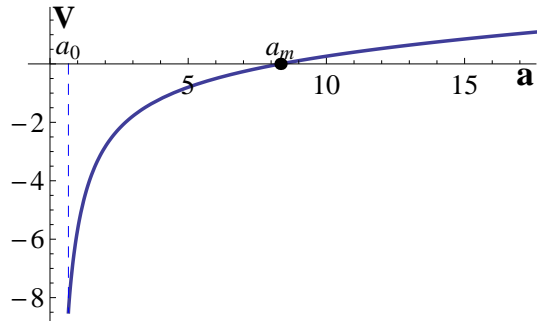}
 \caption{\label{F11} The potential $V(a)$ in a matter dominated universe for the case of $\epsilon=-1$ and $k=-1$. The constants are  $\epsilon=-1$, $k=-1$, $\eta=1$, $\mu=1$, $\rho_0=0.1$, $\mathcal{E}_0=10$, and $g_m=15$. }
\end{figure}

From Fig.~(\ref{F11}) we find that, when $a_{0}\leq a\leq a_m$,
$V\leq0$. So, a $FS\Rightarrow FS$ type is obtained.

\section{The evolution of a radiation-dominated universe in the DGP braneworld}

In this section, we discuss the case where the universe is dominated
by radiation $(\omega=\frac{1}{3})$. Thus, the cosmic energy density
can be expressed as $\rho=\frac{g_r}{a^4}$, and the potential
becomes
\begin{eqnarray}\label{var}
V(a)=\bigg(k-\frac{\epsilon}{3\mu^2}\sqrt{2\rho_0(g_r-\mu^2\mathcal{E}_0)}\bigg)-\frac{1}{3\mu^2}\frac{g_r}{a^2}\;.
\end{eqnarray}
It is easy to see that $g_r\geq\mu^2\mathcal{E}_0$ is required. As
in the previous section,  we divide our discussions into two cases:
$\epsilon=+1$, and $\epsilon=-1$.

\subsection{$\epsilon=+1$}

\subsubsection{$k=1$}

\begin{figure}[htbp]
\includegraphics[width=7cm]{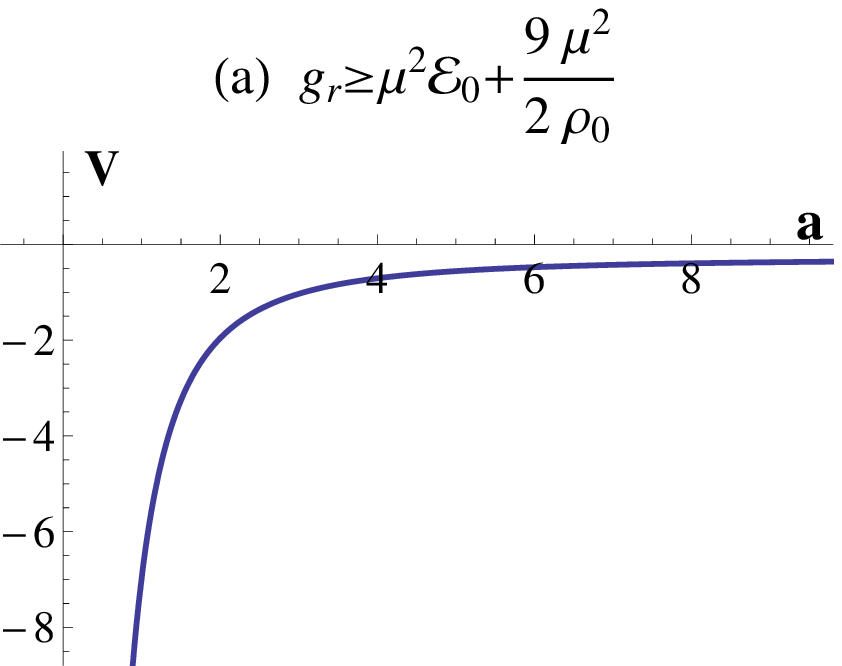}\quad\includegraphics[width=7cm]{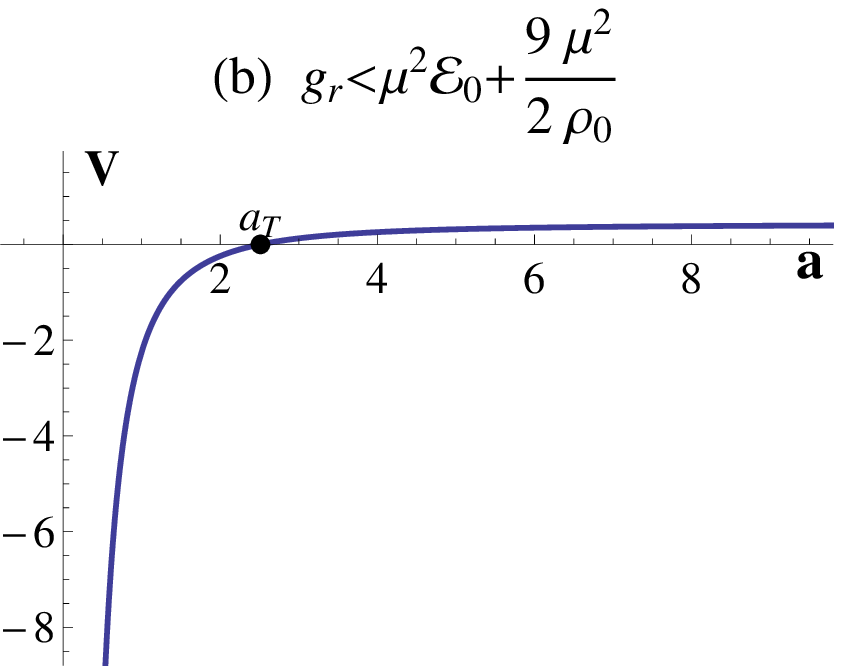}
\caption{\label{Fr1} The potential ${V}(a)$ in a radiation dominated universe  for $\epsilon=1$ and $k=1$ with $g_r\geq\frac{9\mu^2}{2\rho_0}+\mu^2\mathcal{E}_0$
 (left) and $g_r<\frac{9\mu^2}{2\rho_0}+\mu^2\mathcal{E}_0$ (right). The constants are chosen as $\mu=1$, $\rho_0=0.5$, $\mathcal{E}_0=5$, $g_r=20$ (left) and $\mu=1$, $\rho_0=0.5$, $\mathcal{E}_0=5$, $g_r=8$ (right). The radius where the universe turns around is $a_T=2.51185$.}
\end{figure}

As shown in Fig.~(\ref{Fr1}), when
$g_r\geq\frac{9\mu^2}{2\rho_0}+\mu^2\mathcal{E}_0$, the potential is
always negative and the type of the cosmic evolution is
$BB\Rightarrow\infty$ or $\infty\Rightarrow BC$. While, for
$\mu^2\mathcal{E}_0\leq
g_r<\frac{9\mu^2}{2\rho_0}+\mu^2\mathcal{E}_0$, the potential will
turn to be positive from negative at the radius $a_T$
\begin{eqnarray}
a_T=\sqrt{\frac{g_r}{3\mu^2-\sqrt{2\rho_0(g_r-\mu^2\mathcal{E}_0)}}}\;.
\end{eqnarray}
Thus, the cosmic evolution type is $BB\Rightarrow BC$.

\subsubsection{$k=-1$}

\begin{figure}[htbp]
\includegraphics[width=7cm]{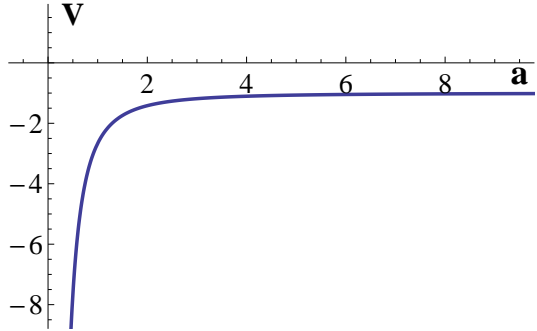}
\caption{\label{Fr2} The potential ${V}(a)$ in a radiation dominated universe for $\epsilon=1$ and $k=-1$. The constants
are chosen as $\mu=1$, $\rho_0=0.5$, $g_r=5$, and $\mathcal{E}_0=5$.}
\end{figure}
From Fig.~(\ref{Fr2}) we can see that in this case the potential is
always negative and the cosmic evolution type is
$BB\Rightarrow\infty$ or $\infty\Rightarrow BC$.

\subsection{$\epsilon=-1$}

\subsubsection{$k=1$}

\begin{figure}[htbp]
\includegraphics[width=7cm]{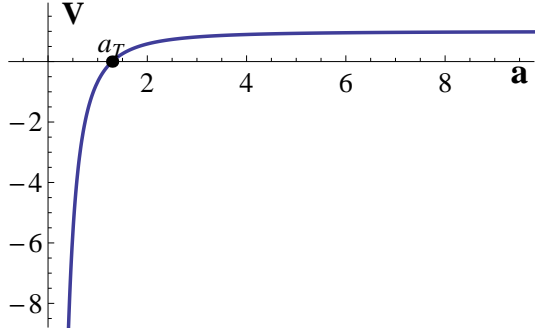}
\caption{\label{Fr3} The potential ${V}(a)$ in a radiation dominated universe for $\epsilon=-1$ and $k=1$. The constants
are chosen as $\mu=1$, $\rho_0=0.5$, $g_r=5$, and $\mathcal{E}_0=5$. The bounce radius is $a_T=1.29099$.}
\end{figure}
 The type of the cosmic evolution now is $BB\Rightarrow BC$ as can be seen from
Fig.~(\ref{Fr3}). The turning point is
\begin{eqnarray}
a_T=\sqrt{\frac{g_r}{3\mu^2+\sqrt{2\rho_0(g_r-\mu^2\mathcal{E}_0)}}}\;.
\end{eqnarray}

\subsubsection{$k=-1$}

\begin{figure}[htbp]
\includegraphics[width=7cm]{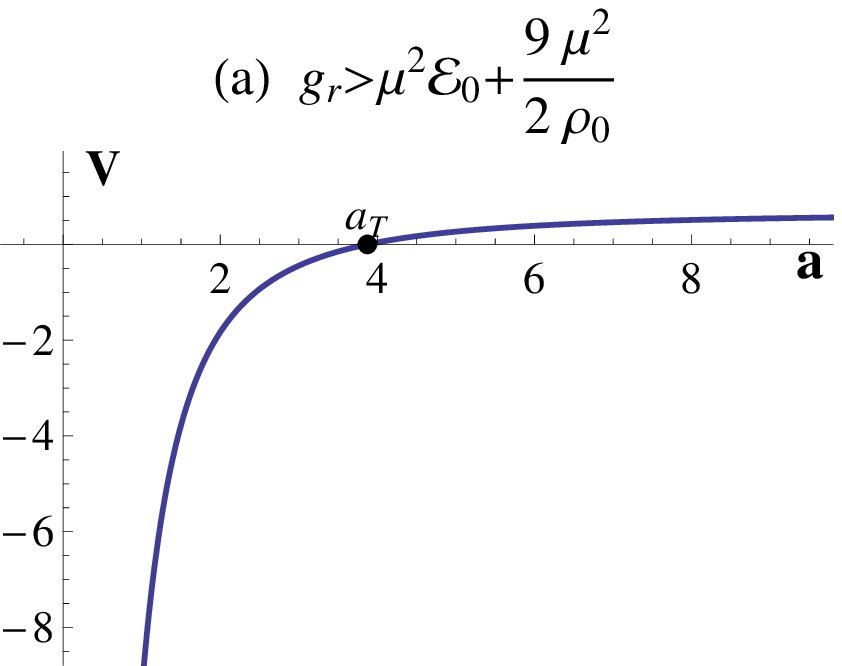}\quad\includegraphics[width=7cm]{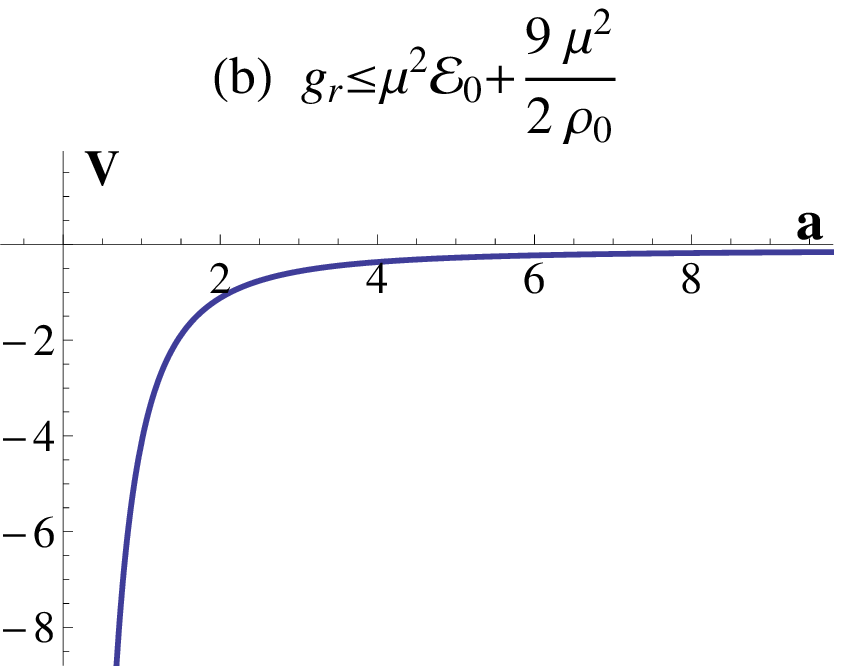}
\caption{\label{Fr4} The potential ${V}(a)$ in a radiation dominated universe for $\epsilon=-1$ and $k=-1$ with $g_r>\frac{9\mu^2}{2\rho_0}+\mu^2\mathcal{E}_0$
 (left) and $g_r\leq \frac{9\mu^2}{2\rho_0}+\mu^2\mathcal{E}_0$ (right). The constants are chosen as $\mu=1$, $\rho_0=0.5$, $\mathcal{E}_0=5$, $g_r=20$ (left) and $\mu=1$, $\rho_0=0.5$, $\mathcal{E}_0=5$, $g_r=8$ (right). The bounce radius is $a_T=3.87298$.}
\end{figure}

From Fig.~(\ref{Fr4}) we obtain that when $g_r>\frac{9\mu^2}{2\rho_0}+\mu^2\mathcal{E}_0$ the cosmic evolution type is $BB\Rightarrow BC$. The radius where the universe turns to contract is
\begin{eqnarray}
a_T=\sqrt{\frac{g_r}{-3\mu^2+\sqrt{2\rho_0(g_r-\mu^2\mathcal{E}_0)}}}\;.
\end{eqnarray}
When $\mu^2\mathcal{E}_0\leq g_r\leq \frac{9\mu^2}{2\rho_0}+\mu^2\mathcal{E}_0$, the cosmic evolution type is $BB\Rightarrow\infty$ or $\infty\Rightarrow BC$ since the potential is always negative.

\section{The evolution of a vacuum-dominated universe in the DGP braneworld}

If the universe is dominated by vacuum energy $(\omega=-1)$, the cosmic energy density is a constant. We denote it by $\rho=g_v$, and the potential becomes
\begin{eqnarray}\label{vav}
V(a)=k-\frac{1}{3\mu^2}g_va^2-\frac{\epsilon}{3\mu^2}\sqrt{2\rho_0(g_va^4-\mu^2\mathcal{E}_0)}\;.
\end{eqnarray}
Clearly, $a\geq a_0=\sqrt[4]{\frac{\mu^2\mathcal{E}_0}{g_v}}$ is needed, and usually $V(a_0)\neq0$ except for the case $k=1$, $\epsilon=-1$ and \begin{eqnarray}\label{a0v}
g_v=\frac{9\mu^2}{\mathcal{E}_0}\;.\end{eqnarray}

\subsection{$\epsilon=+1$}

\subsubsection{$k=1$}

\begin{figure}[htbp]
\includegraphics[width=7cm]{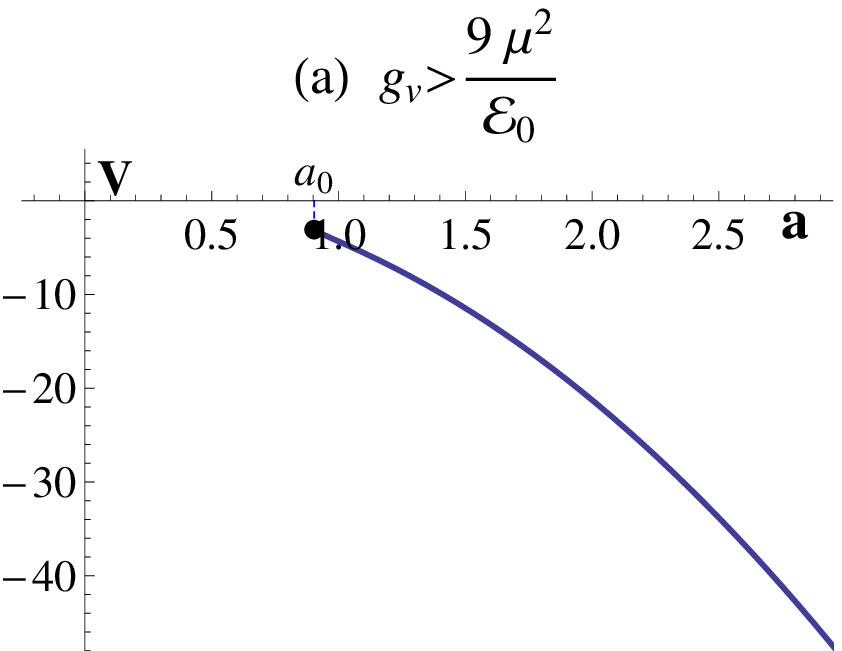}\quad\includegraphics[width=7cm]{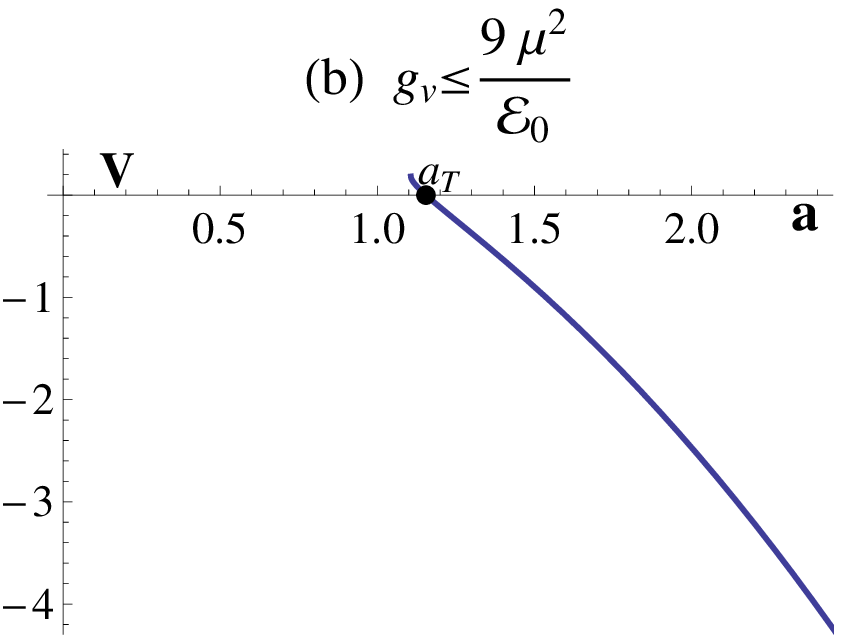}
\caption{\label{Fv1} The potential ${V}(a)$ in a vacuum dominated universe for $\epsilon=1$ and $k=1$ with $g_v>\frac{9\mu^2}{\mathcal{E}_0}$
 (left) and $g_v\leq\frac{9\mu^2}{\mathcal{E}_0}$ (right). The constants are chosen as $\mu=1$, $\rho_0=0.1$, $\mathcal{E}_0=10$, $g_v=15$ (left) and $\mu=1$, $\rho_0=0.1$, $\mathcal{E}_0=3$, $g_v=2$ (right). And the radius where the universe turns around is $a_T=1.1547$. }
\end{figure}
As plotted in Fig.~(\ref{Fv1}), the potential $V(a)$ is a decreasing
function of $a$. If $g_v>\frac{9\mu^2}{\mathcal{E}_0}$, the
potential $V(a\geq a_0)$ is always negative and the type of the
cosmic evolution is $FS\Rightarrow \infty$.  If
$g_v\leq\frac{9\mu^2}{\mathcal{E}_0}$, $V(a_T)=0$, thus, we get a
bouncing universe. The bounce radius is
\begin{eqnarray}
a_T=\sqrt{\frac{3g_v\mu^2-\sqrt{2g_v\rho_0\mu^2(9\mu^2-\mathcal{E}_0g_v+2\mathcal{E}_0\rho_0)}}{g_v(g_v-2\rho_o)}}\;.
\end{eqnarray}

\subsubsection{$k=-1$}

\begin{figure}[htbp]
\includegraphics[width=7cm]{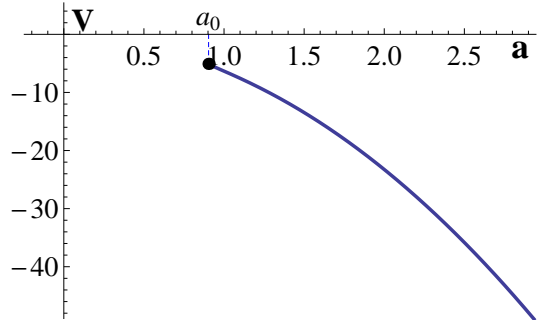}
\caption{\label{Fv2} The potential ${V}(a)$ in a vacuum dominated universe  for $\epsilon=1$ and $k=-1$.  The constants are
chosen as $\mu=1$, $\rho_0=0.1$, $g_v=15$, and
$\mathcal{E}_0=10$.}
\end{figure}

We find from Fig.~(\ref{Fv2}) that the cosmic evolution type is
$FS\Rightarrow \infty$.

\subsection{$\epsilon=-1$}

\subsubsection{$k=1$}

In this case, there exists a static universe if $g_v$ and other parameters satisfy  a relation
\begin{eqnarray}\label{gv}
g_v=\frac{9\mu^2}{\mathcal{E}_0}+2\rho_0=g_v^S\;,
\end{eqnarray}
which is obtained by combining $V(a)=0$ and $V'(a)=0$.  Using the
above equation, one can obtain a static state solution
\begin{eqnarray}\label{as2}
a_S=\sqrt{\frac{\mathcal{E}_0}{3}}\;.
\end{eqnarray}

\begin{figure}[htbp]
\includegraphics[width=7cm]{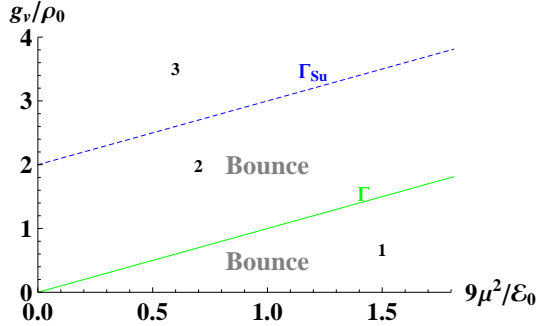}
 \caption{\label{Fv3} Phase diagram of spacetimes in  $(\frac{g_v}{\rho_0},\frac{9\mu^2}{\mathcal{E}_0})$ plane for a vacuum dominated universe with $\epsilon=-1$ and $k=1$. A bouncing universe is found in Regions 1, and
2. $a_0\Rightarrow \infty$ type of universe are found in Region 3. The unstable and stable static universes exist on the curves
$\Gamma_{Su}$ and $\Gamma$, respectively.}
\end{figure}

\begin{figure}[htbp]
\includegraphics[width=7cm]{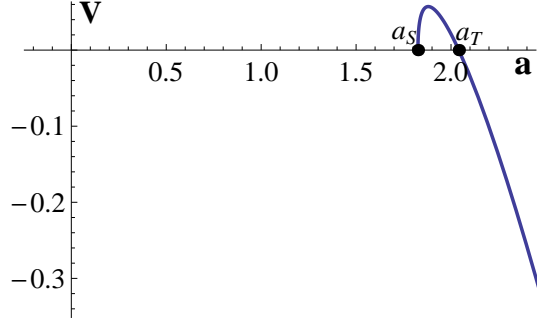}
\caption{\label{Fv4} The potential ${V}(a)$ for a stable static
universe or a bouncing one in a vacuum dominated phase  (the green line of Fig.~(\ref{Fv3})).  The constants are
chosen as $\mu=1$, $\rho_0=0.05$, $g_v=0.9$, and
$\mathcal{E}_0=10$. The radii  are $a_{S}=1.82574$ and $a_T=2.04124$.}
\end{figure}

Using Eqs.~(\ref{a0v}, \ref{as2}), we can depict all cosmic
evolution types in the
$(\frac{g_v}{\rho_0},\frac{9\mu^2}{\mathcal{E}_0})$ plane, which is
shown in Fig.~(\ref{Fv3}).
 On the green line in Fig.~(\ref{Fv3}), which is determined by Eq.~(\ref{a0v}),  we  can get a stable static universe with $a_S=a_0$ and a bouncing
 one as shown in Fig.~(\ref{Fv4}). Thus, the universe can originate
from a stable Einstein static state, which means that  the universe
stays at this stable state past-eternally and then  enters  a
expanding phase through quantum tunneling. If the universe contracts
initially from an  infinite scale, it will bounce at $a_T$. The
radius of the stable static universe is
\begin{eqnarray}\label{aS0}
a_S=a_0=\sqrt{\frac{\mathcal{E}_0}{3}},
\end{eqnarray}
and the bounce radius  is
\begin{eqnarray}\label{aT0}
a_T=\sqrt{\frac{\mathcal{E}_0}{3}+\frac{4\mathcal{E}_0^2\rho_0}{3(9\mu^2-2\mathcal{E}_0\rho_0)}}.
\end{eqnarray}

\begin{figure}[htbp]
\includegraphics[width=7cm]{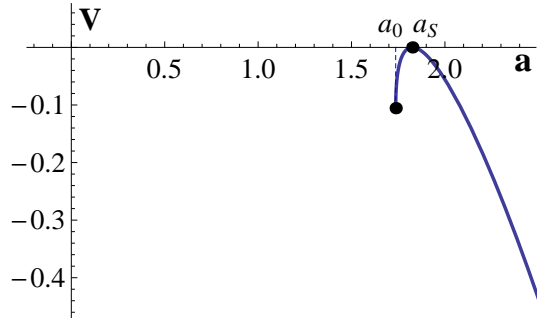}
\caption{\label{Fv5} The potential ${V}(a)$ for an unstable static
universe in a vacuum dominated phase (the blue line of Fig.~(\ref{Fv3})).  The constants are
chosen as $\mu=1$, $\rho_0=0.1$, $g_v=1.1$, and
$\mathcal{E}_0=10$. The radii are $a_{0}=1.73640$ and $a_S=1.82574$.}
\end{figure}

While,  on the blue line of Fig.~(\ref{Fv3}), which is determined by Eq.~(\ref{gv}),  we obtain an unstable static universe. Its radius is
\begin{eqnarray}\label{aSu0}
a_S=\sqrt{\frac{\mathcal{E}_0}{3}}\;.
\end{eqnarray}
We plot the effective potential  in Fig.~(\ref{Fv5}). From which,
one can see that the cosmic evolution type is similar to that shown
in the right panel of Fig.~(\ref{F6}).

\begin{figure}[htbp]
\includegraphics[width=7cm]{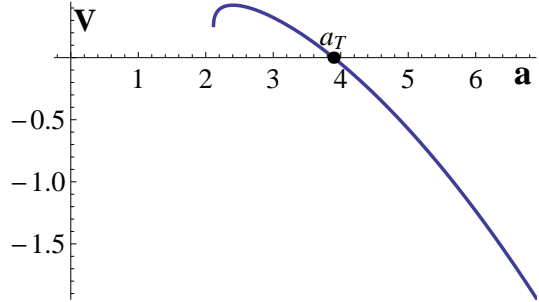}
\caption{\label{Fv6} The potential ${V}(a)$ in a vacuum dominated universe with
the model parameters in Region 1 of Fig.~(\ref{Fv3}). The constants are
chosen as $\mu=1$, $\rho_0=0.1$, $g_v=0.5$. The bounce radius  is $a_T=3.8941$.}
\end{figure}
Fig.~(\ref{Fv6}) shows the evolution of the potential ${V}(a)$ with
the model parameters in  Region 1 of Fig.~(\ref{Fv3}). We find that
a bouncing universe is obtained and  the bounce radius  is
\begin{eqnarray}
a_T=\sqrt{\frac{3g_v\mu^2+\sqrt{2g_v\rho_0\mu^2(9\mu^2-\mathcal{E}_0g_v+2\mathcal{E}_0\rho_0)}}{g_v(g_v-2\rho_o)}}\;.
\end{eqnarray}

\begin{figure}[htbp]
\includegraphics[width=7cm]{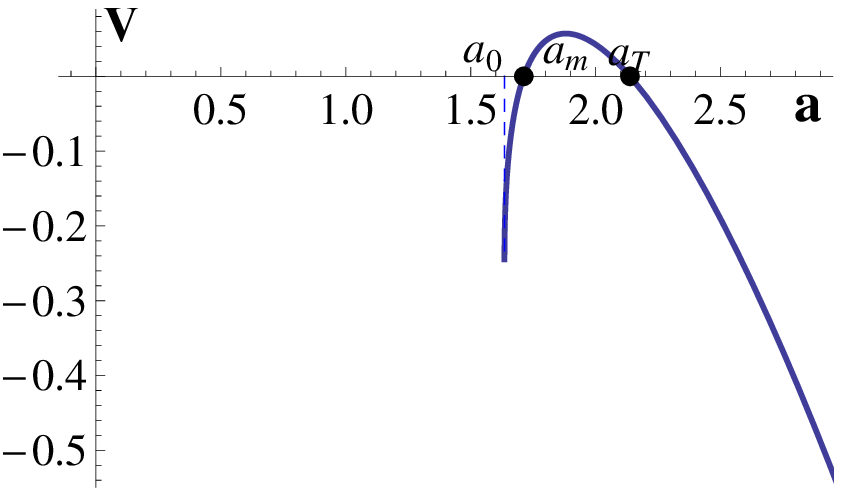}
\caption{\label{Fv7} The potential ${V}(a)$ in a vacuum dominated universe with
the model parameters
in Region 2 of Fig.~(\ref{Fv3}). The constants
are chosen as $\mu=1$, $\rho_0=0.3$, $g_v=1.4$, and $\mathcal{E}_0=10$. The radii are $a_0=1.63481$, $a_m=1.71222$, and $a_T=2.13736$.}
\end{figure}

In Fig.~(\ref{Fv7}), we plot  the potential ${V}(a)$ with
the model parameters in  Region 2 of
Fig.~(\ref{Fv3}). A similar result as shown in Fig.~(\ref{F4})  is obtained.
The expressions for $a_m$ and $a_T$ are
\begin{eqnarray}
a_m=\sqrt{\frac{3g_v\mu^2-\sqrt{2g_v\rho_0\mu^2(9\mu^2-\mathcal{E}_0g_v+2\mathcal{E}_0\rho_0)}}{g_v(g_v-2\rho_o)}}\;,
\end{eqnarray}
\begin{eqnarray}
a_T=\sqrt{\frac{3g_v\mu^2+\sqrt{2g_v\rho_0\mu^2(9\mu^2-\mathcal{E}_0g_v+2\mathcal{E}_0\rho_0)}}{g_v(g_v-2\rho_o)}}\;.
\end{eqnarray}

\begin{figure}[htbp]
\includegraphics[width=7cm]{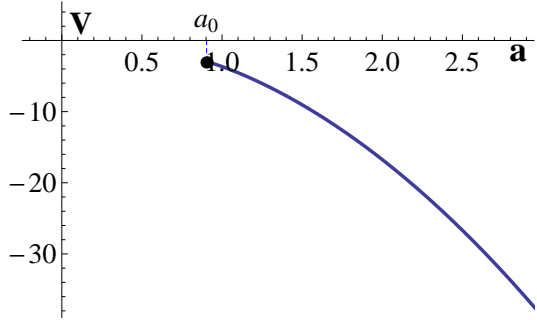}
\caption{\label{Fv8} The potential ${V}(a)$ in a vacuum dominated universe with
the model parameters
in Region 3 of Fig.~(\ref{Fv3}). The constants
are chosen as $\mu=1$, $\rho_0=0.1$, $g_v=15$, and $\mathcal{E}_0=10$.}
\end{figure}
Fig.~(\ref{Fv8}) shows the evolution of the potential ${V}(a)$ with
the model parameters in  Region 3 of Fig.~(\ref{Fv3}), which
corresponds to the cosmic evolution type: $FS\Rightarrow \infty$.

\subsubsection{$k=-1$}

\begin{figure}[htbp]
\includegraphics[width=7cm]{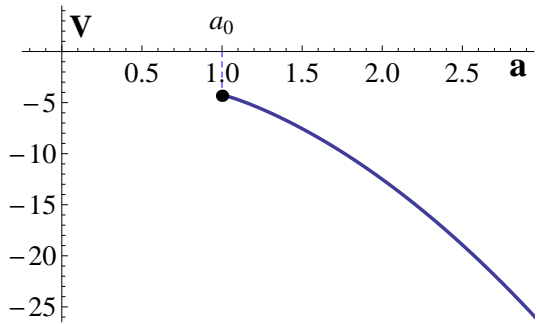}
\caption{\label{Fv9} The potential ${V}(a)$ in a vacuum dominated universe with $\epsilon=-1$ and $k=-1$. The constants
are chosen as $\mu=1$, $\rho_0=0.1$, $g_v=10$, and $\mathcal{E}_0=10$.}
\end{figure}
It is easy to see from Fig.~(\ref{Fv9}) that   $V(a)$ is always
negative. So the cosmic evolution type is $FS\Rightarrow \infty$.

\section{Conclusions}
In this paper, we have studied all possible cosmic evolutions in the
DGP braneworld scenario with a method in which the dynamics of the
scale factor is treated like  that of a particle in a ``potential".
The effect of the dark radiation on the cosmic evolution is
considered. By assuming that the cosmic energy component is
pressureless matter, radiation or vacuum energy, respectively, we
find that, in the matter or vacuum energy dominated case, the
universe does not originate from the big bang singularity and its
scale factor has a minimum value $a_0$. Thus the classical
singularity problem can be avoided. However, there may appear a new
singularity at $a_0$ in the sense that  when the universe bounces or
contracts to this point and then expands, its evolution will be
discontinuous  as $H(a_0)\neq0$. However,  in some circumstances,
there exists a stable Einstein static state or a bouncing universe
to avoid the new and classical singularities. If the universe is in
the Einstein static state initially, it can stay there
past-eternally and evolve to the bounce point through  quantum
tunneling. If the universe contracts initially from an infinite
scale, it can turn around at the bounce point and then expand
forever. Therefore the cosmic evolution is nonsingular. In addition,
in the matter dominated case, there also exists an oscillating
universe to avoid the singularity problem as long as the model
parameters are in some specific regions (shown in Fig.~(\ref{F2})).
If the cosmic scale factor is in the oscillation region initially,
the universe may undergo an oscillation. After a number of
oscillations, it may evolve to the bounce point through  quantum
tunneling. In the radiation dominated case, the universe may
originate from the big bang singularity, but a bouncing universe
which avoids this singularity is also possible.

\acknowledgments  This work was supported by the National Natural
Science Foundation of China under Grants Nos. 10935013, 11175093 and
11075083, Zhejiang Provincial Natural Science Foundation of China
under Grants Nos. Z6100077 and R6110518, the FANEDD under Grant No.
200922, the National Basic Research Program of China under Grant No.
2010CB832803, the NCET under Grant No. 09-0144,  the PCSIRT under
Grant No. IRT0964, the Hunan Provincial Natural Science Foundation
of China under Grant No. 11JJ7001, and the Program for the Key
Discipline in Hunan Province.

%%%%%%%%%%%%%%%%%%%%%%%%%%%%%%%%%%


\begin{thebibliography}{99}
\bibitem{Perlmutter1999}S. Perlmutter, et al., Astrophys. J. {\bf 517}, 565 (1999).
\bibitem{Riess1998}     A. G. Riess,   et al., Astrophys. J. {\bf 116}, 1009 (1998).
\bibitem{Riess2004}     A. G. Riess,  et al.,  Astrophys. J. {\bf 607}, 665 (2004).
\bibitem{Riess2006}     A. G. Riess et al., Astrophys. J. {\bf 659}, 98 (2007).
\bibitem{Astier2006}    P. Astier et al., Astron. Astrophys. {\bf 447}, 31 (2006).  %astro-ph/0510447.
\bibitem{Wood2007}      W. M. Wood-Vasey et al., Astrophys. J. {\bf 666}, 694 (2007).
\bibitem{Dvali2000}    D. Dvali, G. Gabadadze and M. Porrati, Phys. Lett. B {\bf 485}, 208 (2000).
\bibitem{Deffayet2001} C. Deffayet, Phys. Lett. B {\bf 502}, 199 (2001).
\bibitem{Randall1999}  L. Randall, R. Sundrum, Phys. Rev. Lett. {\bf 83}, 4690 (1999).
\bibitem{Shtanov2003} Y. Shtanov, V. Sahni, Phys. Lett. B {\bf 557}, 1 (2003).
\bibitem{Lue2006}A. Lue, Phys. Rept. {\bf 423}, 1 (2006).
\bibitem{Lue2004}      A. Lue, G.D. Starkman, Phys. Rev. D {\bf 70},  101501(R) (2004).
\bibitem{Chimento2006} L.P. Chimento, R. Lazkoz, R. Maartens, I. Quiros, J. Cosmol. Astropart. Phys. {\bf 0609}, 004 (2006).

\bibitem{Bouhmadi-Lopez2004}M. Bouhmadi-Lopez, R. Maartens and D. Wands, Phys. Rev. D {\bf  70}, 123519 (2004).
\bibitem{Cai2004}R. Cai and H. Zhang, J. Cosmol. Astropart. P. {\bf 0408}, 017 (2004).
\bibitem{Papantonopoulos2004} E. Papantonopoulos and V. Zamarias, J. Cosmol. Astropart. P. {\bf 0410}, 001 (2004).
\bibitem{Zhang2004}H. Zhang and R. Cai, J. Cosmol. Astropart. P. {\bf 0408}, 017 (2004).
\bibitem{Zhang2006} H. Zhang and Z. Zhu, Phys. Lett. B {\bf 641}, 405 (2006).
\bibitem{Campo2007} S. del Campo, R.  Herrera, Phys. Lett. B {\bf 653}, 122 (2007).
\bibitem{Zhang}K. Zhang, P. Wu and H. Yu, Phys. Lett. B {\bf 690}, 229 (2010).
\bibitem{Sahni2003} V. Sahni and Y. Shtanov, J. Cosmol. Astropart. Phys. {\bf 0311}, 014 (2003).
\bibitem{Sahni2005} V. Sahni, arXiv: astro-ph/0502032.
\bibitem{Maeda2010}K. Maeda, Y. Misonoh, T. Kobayashi, Phys. Rev. D {\bf82}, 064024 (2010).
\bibitem{Maeda2003}K. Maeda, S. Mizuno and T. Torii, Phys. Rev. D {\bf 68}, 024033
(2003).
\bibitem{Hartle:1983ai}
  J.~B.~Hartle, S.~W.~Hawking,
  %``Wave Function of the Universe,''
  Phys.\ Rev.\ D {\bf28}, 2960-2975 (1983).
\bibitem{Vilenkin:1984wp}
  A.~Vilenkin,
  %``Quantum Creation of Universes,''
  Phys.\ Rev.\ D {\bf30}, 509-511 (1984).

\end{thebibliography}
\end{document}